\documentclass[11pt]{article}

\usepackage{acronym}
\usepackage[ruled,vlined,linesnumbered]{algorithm2e}
\usepackage{amsmath}
\usepackage{amssymb}
\usepackage{authblk}
\usepackage{cleveref}
\usepackage{graphicx}
\usepackage{siunitx}
\usepackage{fullpage}

\newacro{CT}{computed tomography}
\newacro{CTV}{clinical target volume}
\newacro{GTV}{gross tumor volume}
\newacro{MSD}{mean squared displacement}
\newacro{MRI}{magnetic resonance imaging}
\newacro{RT}{radiation therapy}
\newacro{SSA}{stochastic simulation algorithm}

\newcommand{\figbf}[1]{\uppercase{\textbf{#1}}}

\usepackage{xcolor}

\newcommand{\SEAS}{Computational Science and Engineering Laboratory, Harvard John A. Paulson School of Engineering and Applied Sciences, 19 Oxford St, Cambridge, MA 02138, USA}
\newcommand{\MGH}{Massachusetts General Hospital and Harvard Medical School, Department of Radiation Oncology, Division of Radiation Biophysics, 100 Blossom St, Boston, MA 02114, USA}
\newcommand{\Ljubljana}{Faculty of Mathematics and Physics, University of Ljubljana, Ljubljana, Slovenia}
\newcommand{\Freiburg}{University Medical Center, Medical Faculty, University of Freiburg, Germany}

\author[1]{Lucas Amoudruz}
\author[2]{Gregory Buti}
\author[3]{Luciano Rivetti}
\author[2]{Ali Ajdari}
\author[2]{Gregory Sharp}
\author[1]{Petros Koumoutsakos}
\author[4]{Simon Spohn}
\author[4]{Anca L Grosu}
\author[2,4]{Thomas Bortfeld}

\affil[1]{\SEAS}
\affil[2]{\MGH}
\affil[3]{\Ljubljana}
\affil[4]{\Freiburg}

\title{Ising energy model for the stochastic prediction of tumor islets}
\date{\today}

\begin{document}

\maketitle

\begin{abstract}
  A major challenge in diagnosing and treating cancer is the infiltrative growth of tumors into surrounding tissues.
  This microscopic spread of the disease is invisible on most diagnostic imaging modalities and can often only be detected histologically in biopsies.
  The purpose of this paper is to develop a physically based model of tumor spread that captures the histologically observed behavior in terms of seeding small tumor islets in prostate cancer.
  The model is based on three elementary events: a tumor cell can move, duplicate, or die.
  The propensity of each event is given by an Ising-like Hamiltonian that captures correlations between neighboring cells.
  The model parameters were fitted to clinical data obtained from surgical specimens taken from 23 prostate cancer patients.
  The results demonstrate that this straightforward physical model effectively describes the distribution of the size and the number of tumor islets in prostate cancer.
  The simulated tumor islets exhibit a regular, approximately spherical shape, correctly mimicking the shapes observed in histology.
  This is due to the Ising interaction term between neighboring cells acting as a surface tension that gives rise to regularly shaped islets.
  The model addresses the important clinical need of calculating the probability of tumor involvement in specific sub-volumes of the prostate, which is required for radiation treatment planning and other applications.
\end{abstract}

\section{Introduction}

One of the hallmarks of cancer is the invasion of cancer cells into neighboring tissues, which eventually can lead to metastases in other organs~\cite{hanahan2000hallmarks,hanahan2011hallmarks}.
Most cancer-related deaths are caused by metastasis rather than the primary tumor.
Although modern imaging systems can detect gross disease related to the primary tumor, their spatial and contrast resolution is often insufficient to detect microscopic cancer invasion.

In \ac{RT} of cancer, for example, the clinical strategy is therefore to define the \ac{GTV} as the primary tumor mass seen on medical imaging such as \ac{CT} and \ac{MRI}, and to add a geometric extension of the \ac{GTV} to include surrounding tissue at risk of tumor infiltration, called the \ac{CTV}.
The goal of \ac{RT} treatments is then to deliver a high radiation dose to the \ac{CTV} while sparing the surrounding healthy tissue as much as possible.

\begin{figure}
  \centering
  \includegraphics[width=0.4\textwidth]{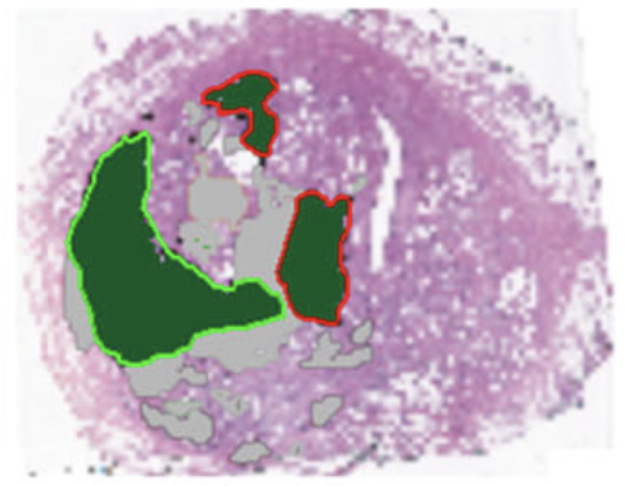}
  \caption{
    Histopathology slice from a prostate cancer patient, adapted from van Houdt et al.~\cite{van2020histopathological} with their permission.
    Green contours indicate the GTV.
    Red contours mark the tumor lesions, invisible with MRI.}
  \label{fig:histopathology}
\end{figure}

Because the tumor cells that infiltrate the tissues surrounding the \ac{GTV} cannot be observed at the time of diagnosis, the \ac{CTV} is either defined as the whole-organ in which the tumor resides (e.g., whole prostate in prostate cancer), or based on population-wide histopathologic studies that analyze the distance of tumorous tissue from the \ac{GTV} in resection specimens.
Currently, simplistic models have been proposed to translate histopathologic data into \ac{CTV} margins.
The most common model is to consider only the distance of the most distant observed tumor for each patient in the population, and subsequently select the \ac{GTV}-to-\ac{CTV} margin as the distance within a certain confidence interval, e.g. 90\% of the population~\cite{Moghaddasi2012}.
However, basing the \ac{CTV} on the most distant tumor observation principle can result in overly conservative treatment plans for many tumor types.
For example, in prostate tumors, histopathologic studies report that infiltrating tumor cells aggregate into small spherical lesions, called tumor foci or \textit{islets}, that are only sparsely distributed around the \ac{CTV}~\cite{johnson2019detection,van2020histopathological,zamboglou2021uncovering,Mouraviev2011}, as illustrated in \cref{fig:histopathology}.
Therefore, treating the \ac{CTV} derived from current models means that most of the dose will be delivered to healthy tissues that do not contain tumor cells at all, potentially unnecessarily compromising the patient's organ function after \ac{RT}.

To take the reduced likelihood of tumor infiltration at increasing distances from the \ac{GTV} into account, numerous recent studies have proposed the definition of the \ac{CTV} as a smooth probability distribution rather than the standard binary \ac{CTV} mask based on \ac{GTV} margins~\cite{Shusharina2018,Bortfeld2021}.
One challenge with this approach is that it often only considers the marginal probabilities of finding tumor cells in each tissue element or voxel.
Neighboring voxels are assumed to be either independent as in~\cite{Shusharina2018}, or fully dependent on its most distant observation as in~\cite{Buti2021}.
This assumption has major implications for designing an optimal radiation dose distribution, as discussed in~\cite{Bortfeld2021,Buti2022}.
Furthermore, in the prostate case mentioned above, the occurrence of tumor in neighboring voxels appears to be only locally correlated: the probability of finding tumor in a voxel is higher if tumor is present in its vicinity.

This work investigates whether a parsimonious model from statistical mechanics, specifically the Ising model, can produce correlated behavior that mimics the spread of tumor cells in prostate cancer patients.

The Ising model, as it is known today, goes back to Wilhelm Lenz's work more than a century ago~\cite{lenz1920beitrag}.
Lenz sought to describe the phenomenon of magnetization in ferromagnetic materials using a lattice of elementary magnets (atomic spins) that could point in only two directions (up or down) and interact only with neighboring spins.
Ernst Ising calculated the partition function for this model in the 1-D case and showed that it does not exhibit spontaneous magnetization, and it therefore does not explain ferromagnetism in 1-D systems~\cite{ising1925beitrag,brush1967history}.

The Ising model has since been generalized to 2-D and 3-D systems, mostly in numerical simulations using computers, which has led to a vast number of successful applications.
It has been applied to describe complex phenomena such as phase transitions not only in physical, but also in chemical and in biological systems. It has even been used as a physical model to address questions of self organization in social networks~\cite{macy2024ising}.

In cancer research, the Ising model has mostly been used at the conceptual level, in game-theoretic cancer modeling approaches~\cite{barradas2018cancer}, and dynamic cancer growth models~\cite{torquato2011toward}.
It has also been used to incorporate the bystander effect in tumor control probability models for radiation therapy~\cite{tempel2018inclusion}.

This study presents a novel application of the Ising model to the problem of tumor infiltration.
The aim is to predict tumor islets of realistic size and shape as reported in the histopathological studies.
The model is applied to prostate cancer patients to estimate the probability of finding islets in a given region around the \ac{GTV}.
We hope that this study will provide a more realistic statistical and quantitative basis for generating the \ac{CTV} in clinical \ac{RT} treatments.

\section{Model}
\label{se:model}

\begin{figure*}
  \centering
  \includegraphics{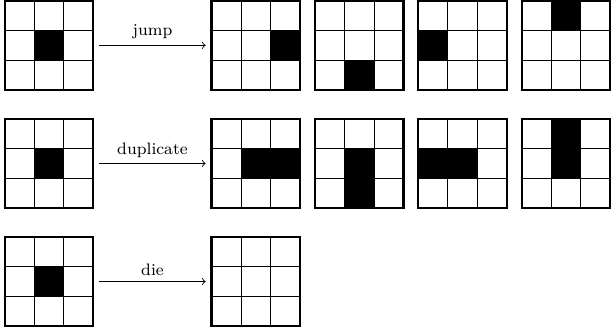}
  \caption{List of possible events for an isolated tumor voxel in two dimensions. Tumor and healthy voxels are represented in black and white, respectively. The left column represent the state before an events occurs, and the configurations on the right are the states after the possible events occur.}
  \label{fig:scheme:events}
\end{figure*}

To model the formation of islets, we discretize the spatial domain into a uniform 3-D grid.
Each grid cell, or voxel, represents a localized region of tissue and can be in one of three discrete states.
\textit{Healthy} voxels correspond to non-tumorous tissues within the organ of interest; \textit{Tumor} voxels indicate the presence of at least one tumor cell; \textit{Boundary} voxels denote regions outside the organ.
The system evolves over time through a set of stochastic events that modify voxel states, each occurring with an associated propensity.
The propensity of an event represents the probability that this event will occur per unit of time.

The basic components of the model are (i) state generation, and (ii) assessing the propensity of events between states.

\subsection{State generation}

We distinguish three types of events that can lead to new states.
We note that these events may not reflect the microscopic dynamics of tumor growth and islets apparition, as these dynamics are not fully understood in-vivo.
The events are represented in \cref{fig:scheme:events}:
\begin{enumerate}
\item \textit{Jump}: A tumor voxel transitions to an adjacent healthy voxel.
  After this event has occurred, the original tumor voxel is converted to a healthy state, and the state of the new location is converted from healthy to tumor.
\item \textit{Duplicate}: A tumor voxel duplicates itself to an adjacent healthy voxel, converting the neighboring voxel from healthy to tumor.
\item \textit{Die}: A tumor voxel converts to a healthy state, representing the disappearance of the tumor at that location.
\end{enumerate}
Neighboring voxels are those that are directly adjacent to the voxel of interest.
In two dimensions, there are 4 neighbors to each voxel, while in three dimensions, each voxel has 6 neighbors.
Therefore, for each tumor voxel, there are up to 4 + 4 + 1 = 9 possible events in two dimensions, and up to 6 + 6 + 1 = 13 possible events in three dimensions.
We remark that jump and duplicate events cannot occur at sites that are marked either as boundary or tumorous.
Each type of event (jump, duplicate, and die) has an associated time scale $\tau_\alpha$, where $\alpha \in \{\text{jump}, \text{dup}, \text{die}\}$.

\subsection{Propensity of events based on the Ising energy}

The propensity associated with each event in the voxel $i$ is given by
\begin{equation}
  a_{\alpha_j,i} = \frac{1}{\tau_{\alpha}} \exp \left(-\frac{\Delta E_{\alpha_j,i}}{k_BT} \right),
\end{equation}
where $k_BT$ is the cells activity, and $\Delta E_{{\alpha_j},i}$ is the change in energy due to the event of type $\alpha$ for the voxel $i$ and neighbor voxel $j$.
The change of energy corresponds to a variation in the Ising model's energy
\begin{equation}
  E = -\frac{1}{2} \sum\limits_{i=1}^N \sum\limits_{\langle j, i\rangle} J \sigma_i \sigma_j,
\end{equation}
where $J$ is the interaction strength, $\langle j, i\rangle$ is the sum over the nearest neighbors $j$ of the voxel $i$ and
\begin{equation}
  \sigma_i =
  \begin{cases}
    +1, & \text{if site i is in tumor state},\\
    -1, & \text{otherwise}.
  \end{cases}
\end{equation}
The Ising model's energy favors the formation of groups of tumor voxels, similar to the islets present in tumor infiltrations~\cite{johnson2019detection,zamboglou2021uncovering}.
Indeed, the change in energy would be positive in the case of a tumor voxel jumping away from the surface of the tumor, resulting in a low propensity associated with such an event.
Equivalently, this energy can be seen as a surface tension term at the surface of tumorous regions, leading to tumor regions with spheroidal shapes.

\subsection{Algorithm for system evolution}

\begin{algorithm}[H]
  \caption{Stochastic Simulation Algorithm (SSA)}
  \label{al:ssa}
  \KwIn{$T_{\text{end}}$: final time; initial voxel grid state}
  \KwOut{Updated voxel states over time}

  $t \gets 0$\;
  \While{$t < T_{\text{end}}$}{
    Compute all propensities $a_i$ for $i = 1, \dots, N_e$\;\label{al:ssa:props}
    Compute total propensity $a_0 = \sum_{i=1}^M a_i$\;
    Draw $r_1, r_2 \sim \mathcal{U}(0,1)$\;
    Compute $\Delta t = \frac{1}{a_0} \ln{\frac{1}{r_1}}$\;
    Select event $\mu$ s.t. $\sum_{j=1}^{\mu -1} a_j < r_2 a_0 \leq \sum_{j=1}^{\mu} a_j$\;\label{al:ssa:cumsums}
    Update $t \gets t + \Delta t$\;
    Update voxels state according to event $\mu$\;
  }
\end{algorithm}


The system evolves according to the \ac{SSA}~\cite{gillespie1976general}, summarized in Algorithm~\ref{al:ssa}, until a specified time horizon $T_\text{end}$.
In \ac{SSA}, events are performed one at a time and are randomly selected according to their propensity.
The time between two events is also randomly sampled and depends on the sum of propensities of all possible events of the system.
After each event, only events associated with nearby sites have a change in their propensities.
Thus, not all operations in line \ref{al:ssa:props} of Algorithm~\ref{al:ssa} need to be performed, and we update only propensities affected by the previous event.
Furthermore, the cumulative sums of propensities in line \ref{al:ssa:cumsums} of Algorithm~\ref{al:ssa}, required for selecting the next event, are maintained using a segment tree.
This data structure enables updates of the cumulative sum in $\mathcal{O}(\log N_e)$ time, compared to $\mathcal{O}(N_e)$ with a naive implementation, where $N_e$ represents the total number of possible events.
These two optimizations significantly reduce the computational cost compared to a naive implementation, since $N_e$ often reaches billions in three-dimensional simulations.

\section{Parameters dependency on grid resolution}

The model presented in Section~\ref{se:model} has four parameters: temperature $k_BT$, and time scales $\tau_\text{jump}$, $\tau_\text{dup}$ and $\tau_\text{die}$.
However, the macroscopic behavior of the model depends not only on the values of these parameters but also on the choice of the grid resolution.
In this section, we relate the model parameters and grid resolution to physical quantities related to the diffusion, growth, and escape rates of tumor voxels.
Let the model parameters for a specific macroscopic behavior be denoted by $k_BT_0$, $\tau_{\text{jump},0}$, $\tau_{\text{duplicate},0}$, and $\tau_{\text{die},0}$ for a reference grid spacing $\Delta x_0$.
We then seek corresponding parameters $k_BT_1$, $\tau_{\text{jump},1}$, $\tau_{\text{duplicate},1}$, and $\tau_{\text{die},1}$ for a new grid spacing $\Delta x_1$ such that the model exhibits the same macroscopic behavior.

First, we observe that \textit{jump} events of isolated tumor cells correspond to a random walk, with time between two jumps following an exponential distribution.
This means that after $n$ steps, the mean squared displacement scales as $n\Delta x^2$, with an average time $n \tau_\text{jump}$.
Thus, the diffusion coefficient associated with this random walk is proportional to
\begin{equation}
  D \propto \frac{\Delta x^2}{\tau_\text{jump}}.
\end{equation}
Therefore, to preserve the same macroscopic diffusion coefficient between the two grid resolutions, the parameter $\tau_\text{jump}$ is adapted as:
\begin{equation} \label{eq:scaling:jump}
  \tau_{\text{jump},1} = \tau_{\text{jump},0} \left(\frac{\Delta x_1}{\Delta x_0}\right)^2.
\end{equation}
Next, we consider the number of tumor vortices that ``escape'' a volume of tumor voxels.
Since we are interested in modeling the formation of islets, the number of tumor vortices that escape this volume per unit of time and for a given surface area $A$ should be independent of the grid spacing.
We therefore require that the following quantity remains constant when varying the grid spacing $\Delta x$:
\begin{equation} \label{eq:scaling:escape}
 \text{escape rate} = \frac{A}{\Delta x^2} \frac{1}{\tau_\text{jump}} \exp\left[ - \frac{\Delta E}{k_BT} \right],
\end{equation}
where $\Delta E$ is the average change of energy resulting from a tumor voxel jumping away from the main tumor region.
Therefore, substituting \cref{eq:scaling:jump} into \cref{eq:scaling:escape} and requiring that the escape rate of voxels due to jump events remains constant as we change the grid spacing, we obtain
\begin{equation} \label{eq:scaling:kBT}
  k_BT_1 = \frac{1}{\frac{1}{k_BT_0} + \frac{4}{\Delta E} \log \frac{\Delta x_0}{\Delta x_1}}.
\end{equation}
Finally, the growth rate of the tumor region, associated with \textit{duplicate} and \textit{die} events, should be independent of grid spacing.
Because of the Ising energy term, we assume that these events mainly happen at the surface of a tumor volume.
Considering a locally flat surface, or low curvature, the region grows perpendicularly to the surface (by symmetry) with a constant velocity
\begin{equation*}
  v \propto \Delta x \alpha_\text{dup},
\end{equation*}
and thus we obtain
\begin{equation} \label{eq:scaling:dup}
  \tau_{\text{dup},1} = \tau_{\text{dup},0} \frac{\Delta x_1}{\Delta x_0} \exp\left[ \frac{\Delta E_d}{k_BT_1} - \frac{\Delta E_d}{k_BT_0} \right],
\end{equation}
where $\Delta E_d$ is the mean change in energy associated with the duplication of a voxel on the surface of the tumor region.
We can repeat the same argument as the growth case for the parameter $\tau_\text{die}$, scaling as
\begin{equation}  \label{eq:scaling:die}
  \tau_{\text{die},1} = \tau_{\text{die},0} \frac{\Delta x_1}{\Delta x_0} \exp\left[ \frac{\Delta E_d}{k_BT_1} - \frac{\Delta E_d}{k_BT_0} \right].
\end{equation}

\begin{figure*}
  \centering
  \includegraphics{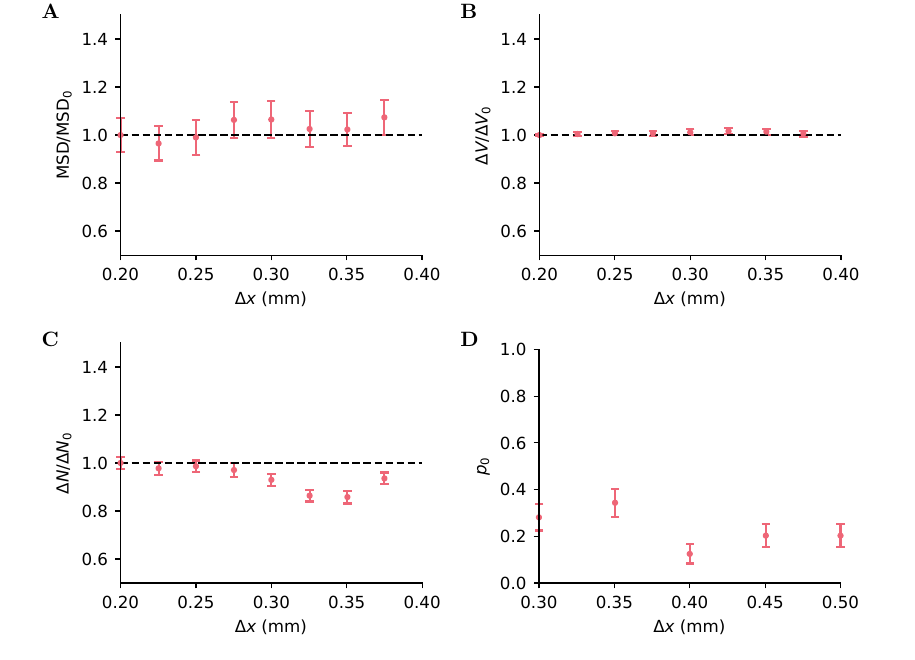}
  \caption{(\figbf{A}) Mean squared displacement (MSD) of a single tumor cell after a fixed simulation time against the grid spacing $\Delta x$. $\mathrm{MSD}_0$ is the MSD obtained with $\Delta x = \SI{0.2}{\milli\meter}$.
    Average over 128 samples, error bars indicate the standard error.
    (\figbf{B}) Volume gain $\Delta V$ after a fixed simulation time against $\Delta x$. $\Delta V_0$ is the mean volume gain obtained with $\Delta x = \SI{0.2}{\milli\meter}$.
    (\figbf{C}) Number of tumor voxels $\Delta N$ that escaped the main tumor region after a fixed simulation time, against $\Delta x$. $\Delta N_0$ is the mean value of $\Delta N$ obtained with $\Delta x = \SI{0.2}{\milli\meter}$.
    Average over 16 samples, error bars indicate the standard error.
    (\figbf{D}) Probability of not finding islets, $p_0$, after a fixed simulation time, against $\Delta x$.
    Average over 64 samples, error bars indicate the standard error.
  }
  \label{fig:conv:dx}
\end{figure*}

First, we consider the case of jump events only.
The grid is initialized with a single tumor voxel and we observe its \ac{MSD} after a fixed time $T=5.0$ with $\tau_\text{jump} = 3.178$ at $\Delta x = \SI{0.3}{\milli\meter}$, $\tau_\text{duplicate} = \tau_\text{die} = \infty$, and $k_BT/J = 0.3167$.
The \ac{MSD} remains constant at all resolutions, within a 10\% margin (\cref{fig:conv:dx}).

Next, we investigate the case of duplicate and die events only.
We consider a periodic domain with large extents of $\SI{150}{\milli\meter}$, and an initial tumor region forming a ball of $\SI{30}{\milli\meter}$ in diameter.
We set $\tau_\text{jump} = \infty$, $\tau_\text{duplicate} = 0.3992$, $\tau_\text{die} = 0.1138$ and $k_BT/J = 0.3167$ at resolution $\Delta x = \SI{0.3}{\milli\meter}$ and let the system evolve until $T_\text{end}=5$.
We then report the change in volume $\Delta V$ of tumor voxels between the initial and final states.
This quantity is shown in \cref{fig:conv:dx} with varying grid spacing and remains within 6\% of the value obtained with the smallest grid spacing.

To test the validity of \cref{eq:scaling:kBT}, we consider the same case as above but with zero propensities associated with duplicate and die events.
We set $\tau_\text{jump} = 3.178$ and $k_BT/J = 0.6$ at $\Delta x = \SI{0.3}{\milli\meter}$ and let the system evolve until $T_\text{end}=5$.
The number of tumor voxels that escape the main tumor region is reported against $\Delta x$ on \cref{fig:conv:dx}.
This quantity converges to a constant as $\Delta x$ decreases.

Finally, we test the convergence of a system with full dynamics.
We use the same domain and initial conditions as in the two previous cases.
Furthermore, we set $k_BT = 0.3167$, $\tau_\text{jump} = 3.178$, $\tau_\text{dup} = 0.3992$, and $\tau_\text{die} = 0.1138$ for a reference grid spacing $\Delta x = \SI{0.3}{\milli\meter}$, and let the system evolve until $T=450$.
At the end of each simulation, islets are counted and we estimate the probability of not finding islets.
Islets are defined as voxel clusters with volume larger or equal to that of a ball of diameter $\SI{1}{\milli\meter}$.
We report the probability of not finding islets, $p_0$, against $\Delta x$ in \cref{fig:conv:dx}.
We observe that, up to statistical noise, the values of $p_0$ remain constant while increasing the grid spacing, indicating that the number of islets produced by the model is independent from the grid spacing when using the scaling rules derived above.

\section{Inference of the model parameters}
\label{se:inference}

In this section, we describe how we fit the parameters of the model described in Section~\ref{se:model}.
The parameters are chosen to match experimental data of Zamboglou et al.~\cite{zamboglou2021uncovering}.
The parameters $J$ and $k_BT$ affect the behavior of the model only in terms of their ratio, $k_BT / J$, thus we choose units such that $J=1$.
Similarly, the parameters $\tau_\text{jump}$, $\tau_\text{dup}$, and $\tau_\text{die}$ have units of inverse time, and thus only their ratio with $T_\text{end}$ affect the output of the model, and we set the simulation time to an arbitrary value, $T_\text{end} = 450$.
We set the grid spacing to $\Delta x = \Delta y = \Delta z = \SI{0.25}{\milli\meter}$.
This resolution corresponds to 4 voxels per diameter of the smallest reported islets~\cite{zamboglou2021uncovering}.

Data from 23 patients' prostates and tumors, extracted from two cohorts, were used to calibrate the model~\cite{bettermann201968ga,zamboglou2021impact}.
This patient data was previously collected as part of a larger study investigating the benefits of PET/CT and multiparametric MRI for detecting prostate cancer~\cite{Zamboglou2017}.
Invasive primary prostate cancer lesions in resected prostates were delineated on histologic tissue slides, matched to ex vivo CT images via a slice-by-slice comparison, and stored in DICOM format.
Plastimatch, open-source software for image computation, was used to convert all DICOM-stored structures into a rasterized grid data format for easy post-processing~\cite{sharp2010plastimatch}.

We extracted lesions from this grid data by computing the connected components between voxels of tumorous type.
Since the largest islets reported in Zamboglou et al.~\cite{zamboglou2021uncovering} was of diameter \SI{5}{\milli\meter}, connected components that were above this threshold were considered as part of the \ac{GTV} and we have used those as initial conditions.
The number of islets and their median diameter were computed from the remaining connected components, and used as the experimental data to match.
Finally, the segmentations of the prostates were used to set boundary voxels in the computational domains.

At the end of each simulation, we compute all connected components formed by the tumor voxels.
Since only islets of diameter larger than $\SI{1}{\milli\meter}$ are reported in ref.~\cite{zamboglou2021uncovering}, we filter out all connected components below a volume corresponding to a sphere with diameter $\SI{1}{\milli\meter}$.
The number of these connected components is then compared with those found in the patients data.
Similarly, we compare the median diameter of the islets predicted by the model with those measured from patients data.
The simulation outputs are stochastic and we thus quantify the number of islets and their median diameters with empirical distributions.
We select the parameters of the model to match these distributions with those computed from the patients data.
To do so, we minimize the Kolmogorov distance between empirical distributions obtained from simulations and experiments, defined as the maximal distance between their cumulative distribution functions,
\begin{equation}
  D_K(p, q) = \max\limits_{x} {| P(x) - Q(x) |},
\end{equation}
where $P$, $Q$ are the cumulative distribution functions of $p$ and $q$, respectively.
The loss function that we minimize is defined as
\begin{equation} \label{eq:loss}
  \mathcal{L}(\theta) = \lambda_n D_k\left(n(\theta), n_\text{exp}\right) + D_k\left( d(\theta), d_\text{exp} \right),
\end{equation}
where $\theta = (k_BT, \tau_\text{jump}, \tau_\text{dup}, \tau_\text{die})$ are the parameters of the model, $n(\theta)$ and $n_\text{exp}$ are the empirical distributions of the number of islets given by the model and reported in experiments, respectively, and $d(\theta)$ and $d_\text{exp}$ the empirical distributions of the median diameters of the islets predicted by the model and reported by experiments, respectively.
The parameter $\lambda_n$ is a positive scalar that weights the importance of matching the number of lesions compared to the median diameters, and was set to $\lambda_n=2$ in this work.
The empirical distributions predicted by the model are estimated from 4 samples for each parameter value $\theta$ and per patient geometry, hence a total of $4\times 23 = 92$ samples per parameter.
\Cref{eq:loss} is minimized using the gradient-free optimization algorithm CMA-ES~\cite{hansen2003reducing} implemented in the software package Korali~\cite{martin2022korali}.
The population size was set to 16 and ran for 9 generations.
The best parameters were found to be $k_BT = 0.450$, $\tau_\text{jump} = 3.726$, $\tau_\text{dup} = 0.149$, and $\tau_\text{die} = 0.103$, and we use these values in the rest of this study.

\begin{figure*}
  \centering
  \includegraphics{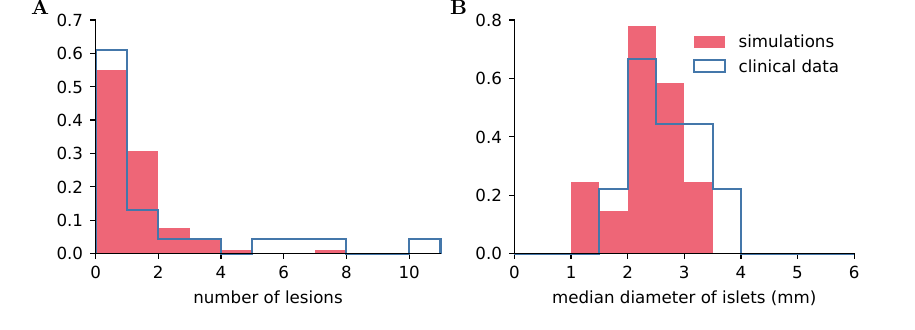}
  \caption{(\figbf{a}) Empirical distribution of the number of lesions found in 23 patients~\cite{zamboglou2021uncovering} and predictions of the model.
    (\figbf{b}) Empirical distributions of the median diameter from ref.~\cite{zamboglou2021uncovering} and predicted by simulations.}
  \label{fig:calibration}
\end{figure*}

A comparison of the empirical distributions of the number of lesions and the median diameters between the simulations and patient data is shown in \cref{fig:calibration}.
The range of values agrees well, between \SI{1}{\milli\meter} and \SI{3.5}{\milli\meter} for the model predictions, while the clinical data shows islets of median diameter ranging between \SI{1.5}{\milli\meter} and \SI{4}{\milli\meter}.
Furthermore, islets with a size around \SI{2.5}{\milli\meter} are the most common in both simulations and clinical data.
This indicates that the model captures the variability observed between patients in a population.
Furthermore, the model predicts that a relatively large portion of the patients (55\%) do not show lesions, as observed in the clinical data (60\%).

\section{Predictions in prostate geometries}

\begin{figure*}
  \centering
  \includegraphics{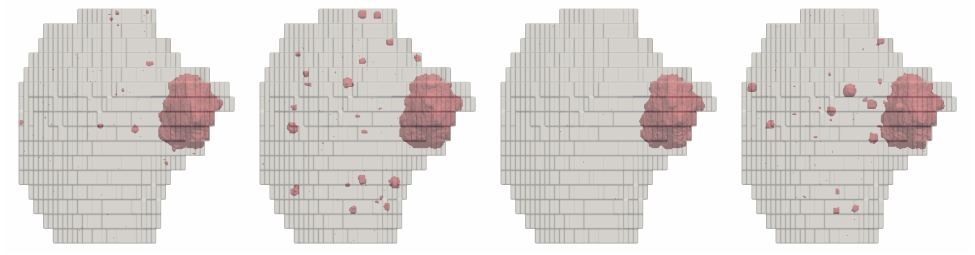}
  \caption{Samples generated with the proposed model in a patient's prostate.
    Red regions indicate tumor voxels at the end of the simulations, and gray area shows the boundaries of the prostate.
    View from the left of the patient.}
  \label{fig:prostate:samples}
\end{figure*}

We now explore the predictions of the model in patient-reconstructed geometries.
In particular, we consider \acp{GTV} inside a prostates reconstructed from MRI scans.
As this imaging technique does not detect all lesions, we use the calibrated model to predict the presence of islets in the prostate.
In the context of radiation therapy, it is crucial to spare vital organs as much as possible.
We are thus interested in the probability of finding tumor cells in the vicinity of nearby organs such as the bladder and the rectum.
More specifically, we want to answer the following question: What is the risk of missing tumor cells if we forego the treatment of a prescribed region near a healthy organ or structure to reduce side effects in these critical structures?
We parameterize the region of interest as the region inside the prostate that is within a distance $d$ from the organ and denote this region $\Omega(d)$.
To estimate the probability of finding islets in this region, we generate $N$ samples with the calibrated model.
The initial conditions correspond to tumor voxels in the \ac{GTV}, boundary voxels are those outside the prostate, and healthy voxels are the remaining voxels inside the prostate.
Each sample $s_k$ corresponds to the final state of a simulation with these initial conditions and a different random seed.
Examples of these samples are shown in \cref{fig:prostate:samples}.
We observe that lesions in the samples have spheroid shapes, as observed in the histological data.
This is due to the inclusion of the Ising energy term in the propensities: Events that would increase this energy have a lower rate than other events, effectively acting as a surface tension.

\begin{figure*}
  \centering
  \includegraphics{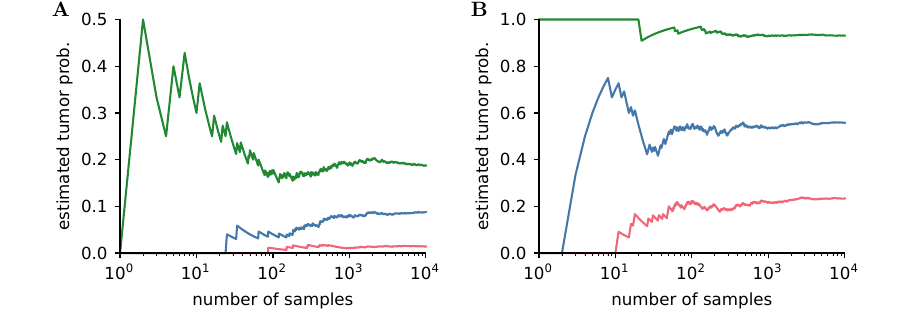}
  \caption{
    Estimate of tumor probability in the regions within a distance $d=\SI{1}{\milli\meter}$, $d=\SI{5}{\milli\meter}$ and $d=\SI{9}{\milli\meter}$ from the bladder (\figbf{a}) and within a distance $d=\SI{0.5}{\milli\meter}$, $d=\SI{0.7}{\milli\meter}$ and $d=\SI{0.9}{\milli\meter}$ from the rectum (\figbf{b}) for patient 2, against number of samples.}
  \label{fig:prostate:convergence}
\end{figure*}

\begin{figure*}
  \centering
  \includegraphics{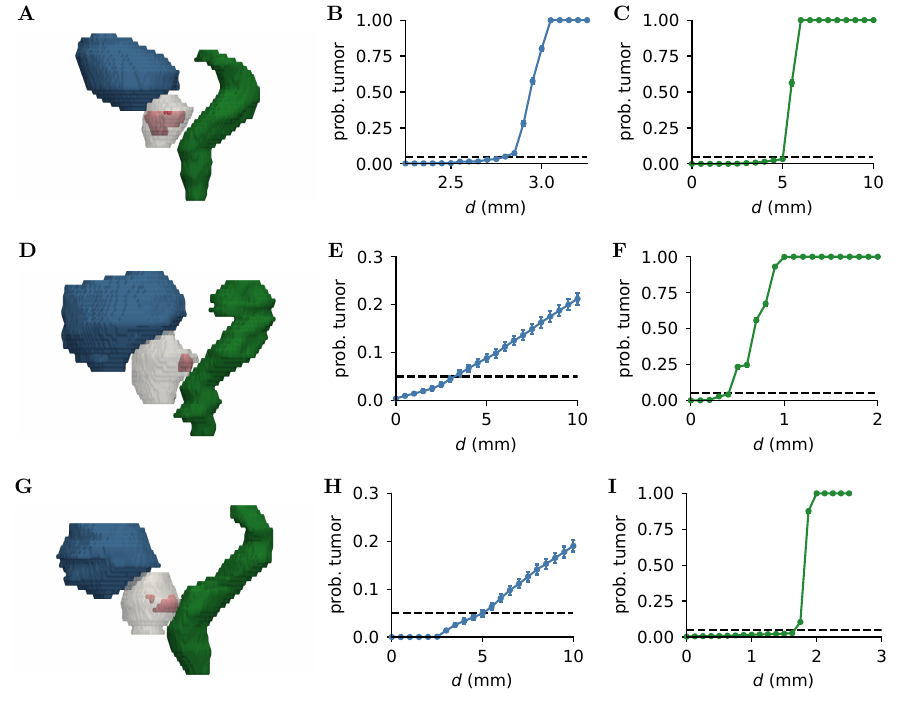}
  \caption{Estimated probability of finding tumor cells in a region within a distance from the bladder (\figbf{b},\figbf{e},\figbf{h}) and from the rectum (\figbf{c},\figbf{f},\figbf{i}) of patients 1, 2, and 3.
    The geometry and initial conditions of the simulations are shown for each patient in (\figbf{a},\figbf{d},\figbf{g}), where the grey region is the prostate, the blue region is the bladder, the green region is the rectum, and the red regions are tumor cells detected by MRI scans, viewed from the left side of the patients.
    Each row corresponds to a different patient.
    Error bars correspond to 3 standard deviations of the estimate given by \cref{eq:estimate:prostate}.
    The horizontal dashed line corresponds to a 5\% probability.
  }
  \label{fig:prostate}
\end{figure*}

For each sample, we record whether the number of tumor voxels within the region of interest exceeds a threshold that corresponds to the volume of a sphere of diameter $\SI{1}{\milli\meter}$.
The probability of finding an islet in this region is then computed with the Monte Carlo estimate
\begin{equation} \label{eq:estimate:prostate}
  P_N(d) = \frac 1 N \sum\limits_{k=1}^N I(s_k, d),
\end{equation}
where $I(s_k, d)=1$ if sample $s_k$ predicted an islet within the region $\Omega(d)$, and $I(s_k, d)=0$ otherwise, and $N$ is the number of samples.
We consider the geometry extracted from three patients, denoted patients 1, 2 and 3, respectively.
The estimate given by \cref{eq:estimate:prostate} is shown in \cref{fig:prostate:convergence} for the bladder and rectum of patient 2 against the number of samples, for three distances $d$ in each case.
In all cases, the estimate converges after a few hundred samples.
The geometry used for this case is shown in \cref{fig:prostate}\figbf{d} together with an estimate of the tumor probability within a distance $d$ from the bladder and the rectum.
We also show these quantities for two other patients.

\Cref{fig:prostate} shows that the tumor probability  against the distance $d$ from both the bladder and the rectum, estimated with $N=10\,000$ samples.
As expected, this value increases monotonically as $d$ increases and vanishes as the region becomes smaller.
This approach allows to find regions where the probability is low, e.g. less than 5\%, to protect vital organs from radiation during therapy.
For example, according to the model, treating the third patient only at a distance greater than $\SI{5}{\milli\meter}$ from the bladder has a low risk of missing tumor cells.
On the other hand, for the second patient, the model suggests that the treatment should include regions that are $\SI{0.5}{\milli\meter}$ from the rectum to remove most tumor cells.
We note that in all patients, the tumor probability close to the rectum increases sharply with the distance $d$.
This is due to the \ac{GTV} being close to the prostate boundary in the vicinity of the rectum, and thus the region $\Omega(d)$ includes the \ac{GTV} before covering a lot of space in the prostate.
We observe a similar behavior for the bladder of the first patient.

\section{Discussion and conclusions}

We have introduced a novel stochastic cellular automaton that models the appearance of tumor islets in prostate cancer. This model can generate islets in realistic geometries extracted from segmented medical images. The state of the system evolves according to three types of events: tumor cells can jump, duplicate, or die. The propensity of an event is lower if this event decreases a nearest-neighbor Ising energy, effectively contributing as a surface tension in the tumor islets. This leads to rounded-shaped islets, consistent with histological observations, and lacking in models available in the literature. We have calibrated the model parameters to match summary statistics extracted from clinical data of 23 patients with prostate cancer. The calibrated model produces empirical distributions of the number and size of islets that agree well with those extracted from clinical data. We note that the parameters were calibrated to match a population of islets using patient-specific geometries. Other patient-specific data, such as the unknown tumor growth rate, were ignored.
Taking these quantities into account may help to predict islets more accurately at the individual level.

\paragraph*{Clinical outlook}

The proposed model can estimate the probability of finding tumor cells in sub-regions of the domain, through a Monte-Carlo approach. Such an estimate can be useful for informing --and ultimately improving-- the design of radiation treatments for cancer patients. In \ac{RT} for prostate cancer, the \ac{CTV} has traditionally been defined as the entire prostate. The concept of more localized treatment options is relatively new. Rather than treating the entire prostate with a uniform radiation dose, clinical trials are investigating whether sparing certain regions can improve the therapeutic window of radiation treatments. One open question is which intraprostatic regions can have the dose escalated or de-escalated. For example, see focal dose escalation in the HypoFocal SBRT trial~\cite{Zamboglou2021b}. Currently, there is a lack of model-based approaches to support dose escalation or de-escalation decisions. We envision the proposed model being clinically applied as a risk model for tumor infiltration at the treatment planning stage. Since the model estimates the probability of finding tumor cells near critical regions, such as the bladder and rectum, it can identify regions where the dose can safely be de-escalated. This would shift the field toward a more evidence-based approach of balancing tumor coverage with sparing healthy tissues. These estimates of tumor presence could be provided without histopathological analysis of all or part of the prostate gland. This could potentially reduce the need for radical or partial prostatectomies and minimize the negative side effects of radiation treatments.

\section*{Acknowledgments}
This project was supported by the National Cancer Institute of the United States under grant number R01CA266275. L. Rivetti received funding from the European Union’s Horizon 2020 Marie Sklodowska-Curie Actions under Grant Agreement No. 955956. We wish to acknowledge the Departments of Urology, Pathology, Nuclear Medicine, and Radiology at the University Clinic Freiburg, Germany, for their support of the data collection and analysis, without which this study would not have been possible. The computations in this paper were run on the FASRC Cannon cluster supported by the FAS Division of Science Research Computing Group at Harvard University.

\bibliographystyle{unsrt}
\bibliography{refs}
\end{document}